\documentclass[prl,twocolumn,showpacs,showkeys,preprintnumbers,amsmath,amssymb,superscriptaddress]{revtex4}

\usepackage{graphicx}
\usepackage{dcolumn}
\usepackage{bm}
\usepackage{units}
\usepackage{color} 
\usepackage[normalem]{ulem}

\newcommand{\ket}[1]{\ensuremath{\left|#1\right\rangle}}

\newcommand{\smm}[1]{\textcolor{black}{#1}}

\begin{document}

\title{Probing individual tunneling fluctuators with coherently controlled tunneling systems}


\author{Saskia M. Mei\ss ner}
\affiliation{Physikalisches Institut, Karlsruher Institut f\"ur Technologie (KIT),
D-76128 Karlsruhe, Germany}

\author{Arnold Seiler}
\affiliation{Physikalisches Institut, Karlsruher Institut f\"ur Technologie (KIT),
D-76128 Karlsruhe, Germany}

\author{J\"urgen Lisenfeld}
\affiliation{Physikalisches Institut, Karlsruher Institut f\"ur Technologie (KIT),
D-76128 Karlsruhe, Germany}

\author{Alexey V. Ustinov}
\affiliation{Physikalisches Institut, Karlsruher Institut f\"ur Technologie (KIT),
D-76128 Karlsruhe, Germany}
\affiliation{Russian Quantum Center, National University of Science and Technology MISIS, Moscow 119049, Russia}

\author{Georg Weiss}
\affiliation{Physikalisches Institut, Karlsruher Institut f\"ur Technologie (KIT),
D-76128 Karlsruhe, Germany}

\date{\today}

\begin{abstract}
Josephson junctions made from aluminium and its oxide are the most commonly used functional elements for superconducting circuits and qubits. 
It is generally known that the disordered thin-film AlO$_{\text{x}}$ contains atomic tunneling systems. 
Coherent tunneling systems may couple strongly to a qubit via their electric dipole moment, giving rise to spectral level repulsion. 
In addition, slowly fluctuating tunneling systems are observable when they are located close to coherent ones and distort their potentials. 
This interaction causes telegraphic switching of the coherent tunneling systems' energy splitting.
Here, we measure such switching induced by individual fluctuators on time scales from hours to minutes using a superconducting qubit as a detector. 
Moreover, we extend the range of measurable switching times to millisecond scales by employing a highly sensitive single-photon qubit swap spectroscopy and statistical analysis of the measured qubit states. 
\end{abstract}

\maketitle

Tunneling systems (TS) are well known to govern the low-temperature properties of glasses, and a quite generally accepted description is provided by the standard tunneling model ~\cite{Phillips:TSAmorphousSolids,Anderson:spin-glasses}.
TS are modeled as two-state systems created by atoms or small groups of atoms residing in double-well potentials. Sufficient overlap of the two localized wave functions results in two coherent states across the two wells. 
Their energy splitting is $E=\sqrt{\varepsilon^2+\varDelta^2}$ with the asymmetry $\varepsilon$ and the tunneling energy $\varDelta$. Interaction with the environment is established via variation of the asymmetry energy $\varepsilon$ by elastic and electric fields.

{\it Coherent} TS can thus be driven resonantly by high frequency elastic or electric fields between their ground and excited states which respectively correspond to symmetric and antisymmetric superpositions of the two localized wave functions. 
{\it Incoherent} TS or so-called {\it two-level fluctuators} (TLF) may be defined as being 
essentially localized in either potential well with a rather low probability of tunneling to the other well. 
The phase of the wave functions is destroyed between subsequent tunneling events ~\cite{Wuerger:CrossOverIncoh,Egger:crossoverCoherentIncoherent}. 
The resulting random telegraph-like occupation of the two positions exerts strain or electric field fluctuations of the local environment which in turn may change the properties of nearby other TSs. 

Here, we employ a phase qubit ~\cite{Steffen:PhaseQbitJL} consisting of a capacitively shunted Josephson junction embedded in a superconducting loop to measure resonantly the state of coherent TSs present in the disordered AlO$_x$ barrier of the junction. 
These TSs act as detectors for nearby incoherent TLFs. 
Being subject to the fluctuating local fields they exhibit jumps of their energy splitting through abrupt shifts of $\varepsilon$.  
The qubit energy is tuned by a flux bias and its state is controlled by externally applied 
microwave pulses. Details of the experimental setup are given by Lisenfeld {\it et al.} 
~\cite{Lisenfeld:TemperatureTS,Grigorij:strainTuningTS}.

Additionally, the energy of both TS and TLF can be tuned by a static strain field which we created by bending the chip with a piezo stack \cite{Grigorij:strainTuningTS}. 
Occasionally, the bending induces irreversible or hysteretic changes of the energy of individual TS directly supporting the picture of locally confined TS-TLF interactions \cite{SupplA}.

\begin{figure}[htb]
	\includegraphics[width=\columnwidth]{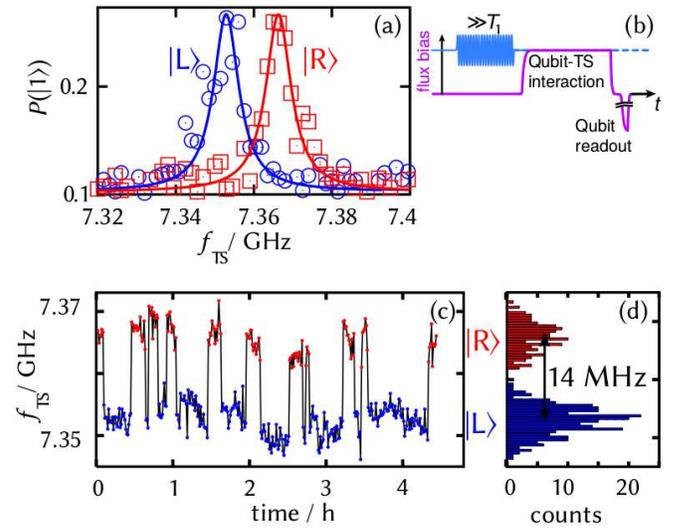}
         \caption{(color online) Energy fluctuation of a TS resonance due to a TLF.
(a) The TS resonances are measured by the qubit. Each data point comprises 1000 
single measurements within $\unit[0.7]{s}$.
(b) Pulse sequence for saturation-spectroscopy.
(c) Random telegraphic noise of the TS resonance measured at $\unit[40]{mK}$. 
(d) Histograms of the population probabilities of the TLF's $|L\rangle$ and $|R\rangle$ states.
	}
	\label{fig:Random-telegraph-1}
\end{figure}
One method to observe slow TLFs with rather long dwell times is to repeatedly measure the resonance curve of an affected TS (Fig.~\ref{fig:Random-telegraph-1}(a)) by exciting it with a long microwave pulse around its resonance (saturation-spectroscopy). 
The excitation of the TS is transferred to the qubit during a swap pulse followed by a qubit state readout (see Fig.~\ref{fig:Random-telegraph-1}(b)).
Figure ~\ref{fig:Random-telegraph-1}(a) shows the resulting probability of the TS to be in the excited state after the long microwave pulse, together with Lorentzian fits, for two frequency sweeps taken a few minutes apart. 
The TS center frequency was extracted from each resonance and plotted as a function of time in Fig.~\ref{fig:Random-telegraph-1}(c), clearly showing telegraphic switching between two frequency values.
A measure of the coupling strength between TS and TLF is given by the difference between the two resonance frequencies and is in this case of about $\unit[14]{MHz}$.
The histograms of the population probabilities of the TLF $|L\rangle$ and $|R\rangle$ states (Fig.~\ref{fig:Random-telegraph-1}(d)) allows one to extract more information about the causative TLF that is not directly visible to the qubit.
Using Boltzmann statistics on the ratio of the dwell times in the localized states of the TLF, one can calculate the energy splitting $E_{\text{TLF}}\approx h\cdot \unit[0.6]{GHz}=k_{\text{B}} \cdot \unit[30]{mK}$ by
\begin{equation}
\frac{\langle\tau_{|R\rangle}\rangle}{\langle\tau_{|L\rangle}\rangle}=e^{-\frac{E_{\text{TLF}}}{k_{\text{B}}T}}
\end{equation}

Changing the static strain by slightly increasing the voltage of the piezo stack results in $E_{\text{TLF}}=k_{\text{B}}\cdot\unit[50]{mK}$ which indicates that $E_{\text{TLF}}$  depends on the asymmetry and is tunable by strain similar to coherent TSs \cite{Grigorij:strainTuningTS}.
The TLF's slow fluctuation rate points towards a small tunneling energy of the order of 
$\varDelta_{\text{TLF}}\approx h \cdot \unit[1]{\mu Hz}$. 
Therefore the energy splitting is given mainly by the asymmetry energy, where 
$E_{\text{TLF}}\approx\varepsilon_{\text{TLF}}\approx k_{\text{B}} \cdot T$ is 
comparable to the temperature $T=\unit[40]{mK}$ of the sample.

The additional small drift of the resonance frequency of the TS in Fig.~\ref{fig:Random-telegraph-1}(c) can be attributed to a larger bath of very weakly coupled TLF and may be discussed in terms of spectral diffusion. 

\begin{figure}[htb]
	\includegraphics[width=\columnwidth]{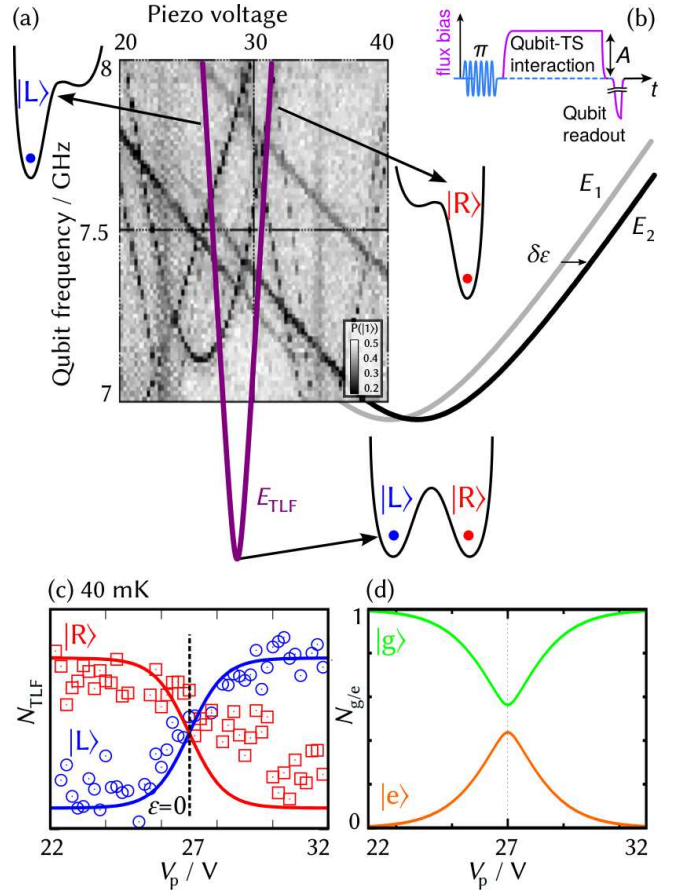}
         \caption{(color online) Strain dependence of TS resonance frequencies: 
         (a) Dark traces correspond to a reduced qubit population indicating that the qubit lost its excitation to a coherent TS. Particular attention is owed to the two parallel TS traces in the middle of the plot with a smooth transition from $E_{1}$ to $E_{2}$ continued as grey and black hyperbolas. The sketches show the suggested TLF energy $E_{\text{TLF}}$ (magenta, energy not to scale) and three variations of its double-well potential. 
         (b) Protocoll for single-photon swap spectroscopy. The qubit is excited by a $\pi$-pulse and subsequently tuned to a range of frequencies to find TSs. 
         (c) Occupation number $N_{\text{TLF}}$ and (d) in energy basis $N_{g/e}$ as a function of strain at $\unit[40]{mK}$ with continous lines obeying equations (\ref{eq:N_TLF}) and (\ref{eq:N_ge}).
         }
	\label{fig:TLF-potential}
\end{figure}
If a TLF switches between its states faster than the time $\tau_M=\unit[0.7]{s}$ required to measure the averaged qubit state probability, both resonance frequencies of the coherent TS appear simultaneously.
Such a situation is depicted in Fig.~\ref{fig:TLF-potential}(a) with a larger overview in \cite{SupplA2}. 
This data was acquired using single-photon swap spectroscopy ~\cite{Lisenfeld:InteractingTS} by applying the pulse sequence shown in Fig.~\ref{fig:TLF-potential}(b).
Dark traces correspond to a reduced probability to measure the excited state of the qubit which indicates that the excitation of the qubit was transferred to a TS.

The resulting change of the TS's hyperbolic trace corresponds to a shift along the strain axis, indicating that the coupling to the TLF only affects the TS's asymmetry and not its tunneling energy.
For the two parallel TS resonances $E_{1}$ and $E_{2}$ (grey and black hyperbolas), 
which are simultaneously visible only in a small range of mechanical deformation (Fig.~\ref{fig:TLF-potential}(b)), 
the suggested potential of the TLF (magenta hyperbola) strongly depends on external strain.
Coming from a highly asymmetric potential configuration where it is trapped most of the time in the left state, the TLF passes through its symmetry point and finally ends up in the right state. Both traces of the resonant TS visible to the qubit are described by the same hyperbola only shifted by a mechanical distortion corresponding to a change of the piezo voltage of $\unit[2.45]{V}$. 
The changes in the density of the two TS traces correspond to the change in the 
left-right occupancy of the TLF.

From the qubit population extracted along the two branches in Fig.~\ref{fig:TLF-potential}(a) we obtain the occupancy number $N_{\text{TLF}}$ of the TLF as a function of mechanical 
deformation (Fig.~\ref{fig:TLF-potential}(c)). 
Due to the fact that there is an intersection of the two traces it becomes clear that the TLF is measured in its localized states in the right or left potential well, described by   

\begin{equation}
N_{\text{TLF}_{\text{L/R}}}=\frac{\varDelta^2}{2\left(E^2\pm\varepsilon E\right)}\left(N_{\text{g}}+\left(\frac{E\pm\varepsilon}{\varDelta}\right)^{2}N_{\text{e}}\right) 
\label{eq:N_TLF}
\end{equation}
where
\begin{equation}
N_{\text{g/e}}=\frac{1}{2}\left(1\pm\tanh\left(\frac{E}{2k_{\text{B}}T}\right)\right) 
\label{eq:N_ge}
\end{equation}
is the occupation number in the energy basis (Fig.~\ref{fig:TLF-potential}(d)).

Although we can not resolve telegraph switching for this rather fast TLF there is a way to extract information about its switching rate using a statistical analysis of many subsequent individual qubit measurements. 
\smm{In principle the data can be analyzed as well in the framework of autocorrelation  \cite{FCS:Elson}. 
Here we prefer to simulate the statistics of the quantum measurement.}

\begin{figure}[htb]
	\includegraphics[width=\columnwidth]{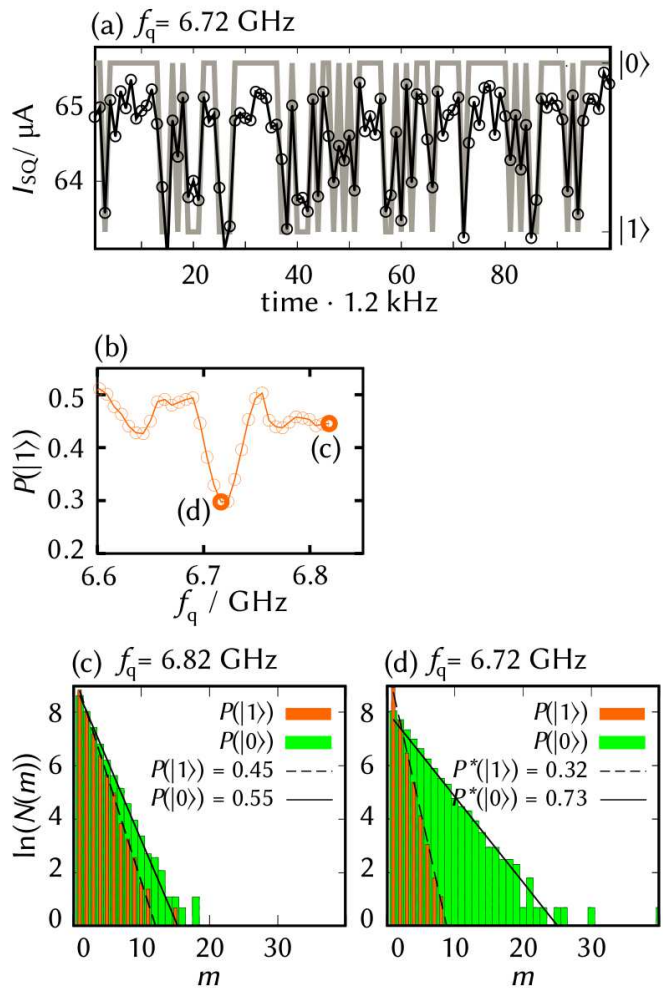}
         \caption{(color online) (a) $100$ out of $50.000$ single measurements of the readout DC-SQUID's switching current $I_{SQ}$ (black line, left axis) and the associated qubit states $\ket{0}$ and $\ket{1}$ (grey line, right axis).
         (b) Qubit population probability $P\left(|1\rangle\right)$ for resonant excitation when it was tuned to different resonance frequencies $f_\mathrm{q}$. The dip at $f_\mathrm{q}=$ 6.72 GHz indicates resonant interaction with a coherent TS. (c) and (d) show the abundancies $\mathrm{ln}(N(m))$ of measuring $m$ successive times the same qubit state $P\left(|0\rangle\right)$ (green) or $P\left(|1\rangle\right)$ (orange), taken either for the isolated qubit (c) or the qubit in resonance with the TS (d). For the latter case, the qubit is found with a larger probability of $P^{\,*}|0\rangle$ in its ground state. Continuous and dashed lines are fits to Eq. (\ref{eq:abundance}). \smm{Deviations from the unity of the sum of probabilities are due to statistical uncertainty using a finite number of measurements.}        
	}
	\label{fig:TS-statistics}
\end{figure}

In the following, we present such a statistical analysis and a corresponding simulation using a sequence of $N_{0}=50.000$ successive single measurements.
Figure~\ref{fig:TS-statistics} (a) shows a series of successive individual measurements of the readout DC-SQUID's switching current which depends on the excitation probability of the qubit. 
By defining a threshold value, we attribute each switching current to one of the two qubit states as shown by the grey digital data and right vertical axis.
Figure~\ref{fig:TS-statistics} (b) shows the qubit excitation probability, where the qubit was biased to different resonance frequencies $f_\mathrm{q}$ averaged from 50.000 measurements for each $f_\mathrm{q}$. 
Due to energy relaxation which occurs at a rate of $T_1^{-1} \approx 1/100$ns during the 40 ns-long qubit-TS interaction time (see Fig. ~\ref{fig:TLF-potential} (b)), the qubit remains at a maximum excitation probability of $P(\ket{1})\approx 0.5$ when it is not in resonance with a strongly coupled TS. In contrast, at a frequency of $f_ {\text{q}} =\unit[6.72]{GHz}$, resonant interaction with a TS results in a reduced excitation probability of $P^{\, *}\left(|1\rangle\right)=0.32$ 
since the energy was transmitted to the TS with a certain probability.

The probabilities $P\left(|1\rangle\right)$ and $P\left(|0\rangle\right)$ to measure the excited or the ground state of the qubit, respectively, are given by a Bernoulli distribution with two possible outcomes and $P\left(|0\rangle\right)+P\left(|1\rangle\right)=1$ \cite{Uspensky:MathPropability}.
One finds that the numerical simulation of our coupled detector system is similar to the statistics problem of tossing a biased coin for which $P\left(|0\rangle\right) \neq P\left(|1\rangle\right)$.
In the case of independent individual measurements a closed expression for the abundance $N(m)$ of exactly $m$ successive measurements which have the same result is described by a Bernoulli distribution \smm{for $m \ll N_0$}
\smm{
\begin{equation}
\ln\left(N\left(m\,\right)\,\right)=m\cdot\ln\left(p\,\right)+\ln\left(N_{0}\,\right)+2\ln\left(1-p\right)
\label{eq:abundance}
\end{equation}
}
where $p$ is either $P\left(|1\rangle\right)$ or $P\left(|0\rangle\right)$ (Supplemental Material \cite{SupplB}).
Therefore the abundance $N(m)$ of measuring $m$ times the same qubit state results in two histograms for the qubit's ground $\ket{0}$ (orange) and excited $\ket{1}$ (green) state which is depicted in Fig. \ref{fig:TS-statistics}(c) for the isolated qubit and (d) for the qubit interacting with one coherent TS. 
Thus, the slope of the straight black lines using Eq. (\ref{eq:abundance}) in a logarithmic plot coincides with the measured probabilities (Fig.~\ref{fig:TS-statistics}) for both analyzed frequencies $f_ {q}$.

This indicates that the statistics for the case of an isolated qubit
and for the case of the qubit in resonance with a TS, are both described by a Bernoulli process, proofing the independence of subsequent events, where however the latter case results in an increased decay probability of the qubit. 

\begin{figure}[htb]
	\includegraphics[width=\columnwidth]{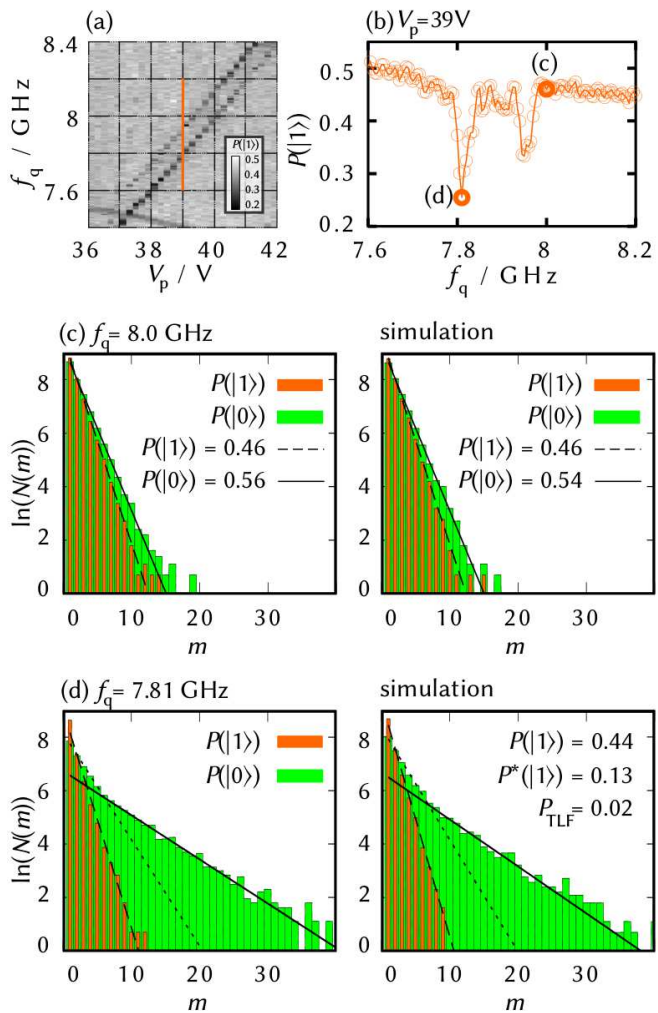}
         \caption{(color online)  
         (a) Two parallel TS traces as a function of mechanical deformation. 
         (b) Line cut along the frequency axis (orange line) of (a) near the symmetry point of the TLF with two highlighted data points marked with (c) and (d) according to the following plots. 
         (c) Statistic analysis using $50.000$ single qubit measurements and simulation of the qubit without pertubation: The abundance $N(m)$ of measuring $m$ times the same qubit state results in two histograms for the qubit's ground $\ket{0}$ (green) and excited $\ket{1}$ state (orange), black dashed lines are fits with Eq. (\ref{eq:abundance}).
         (d) Analysis of a coherent TS which is disturbed by an incoherent TLF: 
         The switching probability $P_{\text{TLF}}$ is determined by comparison of the simulation with the experimental data (black lines are only guides to the eye).
	}
	\label{fig:Dwell-time-TLF}
\end{figure}

Now the statistical analysis is applied to a coupled system consisting of the qubit 
and a coherent TS which is additionally modulated by a TLF. 
In Fig.~\ref{fig:Dwell-time-TLF}(a) another example of two TS resonance traces as a function of mechanical deformation are shown. 
A vertical cut of this plot close to the symmetry point of the TLF is shown in 
Fig.~\ref{fig:Dwell-time-TLF}(b) where the two highlighted data points are analyzed further.
As shown in Fig.~\ref{fig:Dwell-time-TLF}(c), the abundance $N(m)$ of measuring $m$ times the same qubit state results in two histograms for the qubit's ground $\ket{0}$ (orange) and excited $\ket{1}$ (green) states. 
The exponential abundance measured at $f_{q}=\unit[8]{GHz}$ (Fig.~\ref{fig:Dwell-time-TLF}(c)) $N(m)$ agrees with Eq.~(\ref{eq:abundance}) in accordance to the numerical simulation of an undisturbed Bernoulli process. 

Analysis of a TS which is coupled to an incoherent TLF reveals a correlation of subsequent single measurements of the qubit state.
The correlation manifests itself as a kink in the histogram of Fig.~\ref{fig:Dwell-time-TLF}(d) measured at $f_{q}=\unit[7.81]{GHz}$ where the qubit is in resonance with the TS.
The coupled system of the qubit, the TS, and the TLF is simulated by a coin toss with two differently biased coins. 
One coin with a specific probability $P^{\, *}$ describes the qubit and the TS. 
Another coin with $P$ describes the qubit without perturbation, where the TLF has shifted the TS out of resonance with the qubit. 
With a certain probability $P_{\text{TLF}}$, the TLF exchanges the two coins between two single measurements in the series of $N_{0}$ measurements, so that either $P$ or $P^{\, *}$ is measured (Supplemental Material \cite{SupplB}). 
This simulation is used in Fig.~\ref{fig:Dwell-time-TLF}(d) where the switching probability $P_{\text{TLF}}$ of the TLF is determined by adjusting the simulation parameters until agreement with the experimental data was found.
\smm{We want to note that the TLF switching rates from $|R\rangle$ to $|L\rangle$ and vice versa in general are not equal but depend on their energetic difference.} 
However, since the observed TLF is close to its symmetry point ($\varepsilon_{\text{TLF}}\approx 0$), we can treat them as being approximately equal.

The extracted probability $P_{\text{TLF}}=(0.02\pm0.01)$ and the repetition rate of $\unit[1.2]{kHz}$ results with $\tau_{\text{TLF}}^{-1}=\unit[1.2]{kHz}\cdot P_{\text{TLF}}$ in a fluctuation rate of
\begin{equation}
 \tau_{\text{TLF}}^{-1} = 1 / \unit[(38\pm17)]{ms}.
\end{equation}
The significance of this result has to be discussed briefly. 
When comparing the experiment with the coin toss simulation, the observation of the characteristic kink is crucial, and the position of the kink mainly depends on $P_{\text{TLF}}$ \cite{SupplB}.
Clearly, the determination of decay rates with such a (statistical) method has an upper bound given by the repetition rate of the experiment. 
Here, the repetition rate is limited by the initialization and readout of the qubit. 
Faster methods might be possible with dispersive readout protocols that allow repetition rates of $\approx 1 - 10$ MHz \cite{Walter:RapidQBReadout}. 

In conclusion, individual TLFs in the AlO$_x$ of Josephson junction tunnel barriers and their switching dynamics are measured through a two stage detection involving their coupling to a coherent TS which itself is coupled to a qubit.
Various existing methods to characterize slow TLFs, e.g. conductance fluctuations,  are limited to $\approx \unit[1]{Hz}$ by the averaging time of the measuring system \cite{ChunBirge:QMtunnelingBi1993,Brouer:BiMechTS,Yehe:NanoGrainDynamics}. 
Here, we have described a method to measure much faster fluctuation rates, up to 125\,Hz.

Finally we would like to thank J. M. Martinis (UCSB) for providing
us with the sample that we measured in this work,
\smm{as well as D. Hunger and S. Matityahu for fruitful discussions.}
Support by the Deutsche Forschungsgemeinschaft (DFG) (Grant No. LI2446/1-1) is gratefully acknowledged, as well as partial support by the Ministry of Education and Science of Russian Federation in the framework of Increase Competitiveness Program of the NUST MISiS (Grant No. 2-2016-063).


\begin{thebibliography}{11}
\bibitem{Phillips:TSAmorphousSolids} W. A. Phillips, J. Low Temp. Phys. {\bf 7}, (1972).
\bibitem{Anderson:spin-glasses}  P. W. Anderson, B. I. Halperin, and C. M. Varma, Phil. Mag. {\bf 25}, 19 (1972). \url{http://dx.doi.org/10.1080/14786437208229210.}
\bibitem{Wuerger:CrossOverIncoh} A. W\"urger, R. Weis, M. Gaukler, and C. Enss, Europhys. Lett. {\bf 33}, 533 (1996). 
\bibitem{Egger:crossoverCoherentIncoherent} R. Egger, H. Grabert, and U. Weiss, Phys. Rev. E {\bf 55}, R3809R3812 (1997). 
\bibitem{Steffen:PhaseQbitJL} M. Steffen, M. Ansmann, R. McDermott, N. Katz, R.C. Bialczak, E. Lucero, M. Neeley, E.M. Weig, A.N. Cleland, and J.M. Martinis, Phys. Rev. Lett. {\bf 97}, 050502 (2006). 
\bibitem{Lisenfeld:TemperatureTS} J. Lisenfeld, C. M\"uller, J. Cole, P. Bushev, A. Lukashenko, A. Shnirman, and A. V. Ustinov, Phys. Rev. Lett. {\bf 105}, 230504 (2010). 
\bibitem{Grigorij:strainTuningTS} G. J. Grabovskij, T. Peichl, J. Lisenfeld, G. Weiss, and A. V. Ustinov, Science {\bf 338}, 232234 (2012). 
\bibitem{Lisenfeld:InteractingTS} J. Lisenfeld, G. Grabovskij, C. M\"uller, J. H. Cole, G. Weiss, and A. V. Ustinov, Nat. Commun. {\bf 6}, 7182 (2015). 
\bibitem{FCS:Elson} E. L. Elson, Biophys. J. {\bf 101}, 28552870 (2017).
\bibitem{Uspensky:MathPropability} V. Uspensky, {\it Introduction to Mathematical Probability}, McGraw-Hill New York, (1937).
\bibitem{ChunBirge:QMtunnelingBi1993} K. Chun and N. O. Birge. Phys. Rev. B, {\bf 48} (1993).
\bibitem{Brouer:BiMechTS} S. Brou\"er, G. Weiss, and H. B. Weber, Europhys. Lett. {\bf54}, 654 (2001), 
\bibitem{Yehe:NanoGrainDynamics} S.-S. Yeh, W.-Y. Chang, and J.-J. Lin, Science Advances {\bf 3} (2017). 
\bibitem{Walter:RapidQBReadout} T. Walter {\it et al.}, Phys. Rev. Applied {\bf 7}, 054020 (2017).
\bibitem{SupplA} Supplemental Material at [URL will be inserted by publisher] for Local hysteresis of TS due to TLF.
\bibitem{SupplA2} Supplemental Material at [URL will be inserted by publisher] for Fast TLF.
\bibitem{SupplB} Supplemental Material at [URL will be inserted by publisher] for Statistical analysis.
\end{thebibliography}

\clearpage
\newpage
\begin{widetext}
\begin{center}
\textbf{\large Supplemental Materials: Probing individual atomic tunneling fluctuators}
\end{center}
\end{widetext}
\setcounter{equation}{0}
\setcounter{figure}{0}
\setcounter{table}{0}
\setcounter{page}{1}
\makeatletter
\renewcommand{\theequation}{S\arabic{equation}}
\renewcommand{\thefigure}{S\arabic{figure}}
\renewcommand{\bibnumfmt}[1]{[S#1]}
\renewcommand{\citenumfont}[1]{S#1}

\section*{Local hysteresis of TS due to TLF}
\begin{figure}[htb]
	\includegraphics[width=\columnwidth]{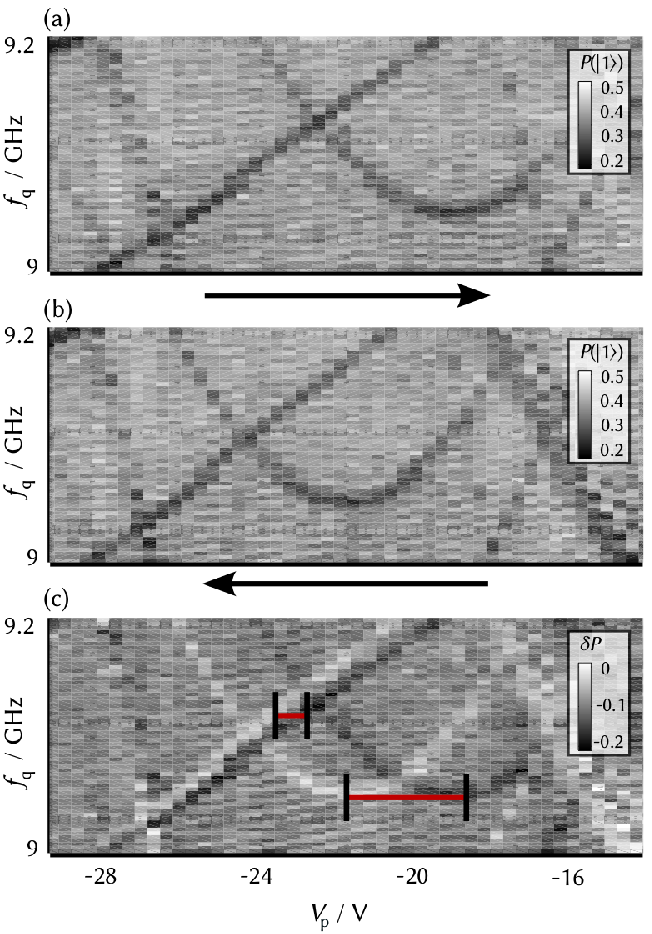}
         \caption{(color online) Local hysteresis of TS energies
         as a function of mechanical distortion in (a) push and (b) release directions. 
         (c) The difference of the data (a) and (b) shows that the TS asymmetry changes by different amounts for different TS (horizontal lines),
         dark: with increasing the voltage; light: while lowering the voltage. 
         Apart from the global hysteresis of the piezo stack, local rearrangements lead to new equilibrium positions of individual TSs.         
	}
	\label{fig:hysteresis}
\end{figure}
A measurement to determine the hysteresis of the piezo stack for the mechanical deformation of the qubit chip shows that, apart from the global hysteresis of the piezo, different TS experience different local hysteresis (Fig.~\ref{fig:hysteresis}). 
The difference between the two measurements in the push (Fig.~\ref{fig:hysteresis} (a)) and release (Fig.~\ref{fig:hysteresis} (b)) of the mechanical deformation shows that certain TS resonance frequencies are not shifted by the same amount (Fig.~\ref{fig:hysteresis} (c)).
An intuitive explanation is given by the local modification of the potential of the coherent TS due to slow TLF.

\section*{Fast TLF}

\begin{figure}[htb]
	\includegraphics[width=\columnwidth]{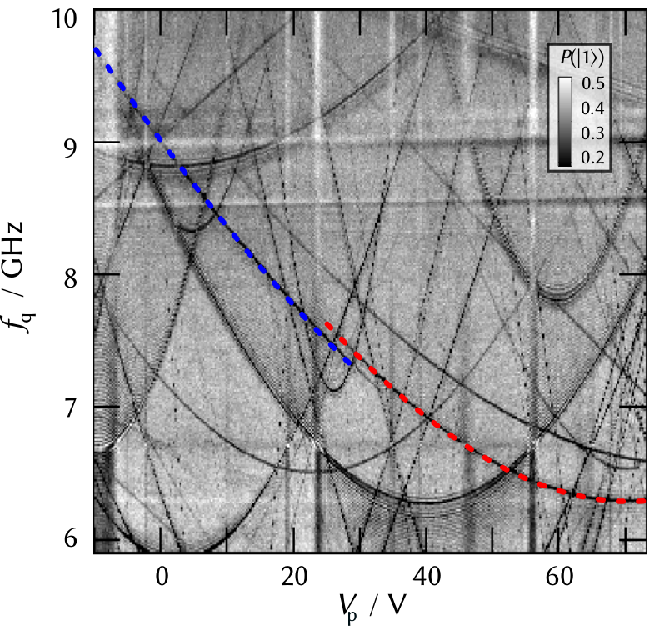}
         \caption{(color online) Strain dependence of TS resonance frequencies: 
         dark lines correspond to a reduced qubit population indicating that the qubit lost its excitation to a coherent TS. Particular attention is owed to the two parallel TS traces in the middle of the plot with a smooth transition from $E_{1}$ (blue dotted line) to $E_{2}$ (red dotted line).
	}
	\label{fig:fastspec}
\end{figure}

\section*{Statistical analysis}

Within the single-photon swap spectroscopy the qubit is excited by a $\pi$-pulse, then it is brought into resonance with a TS for a certain time. 
In this time the excitation oscillates between qubit and TS. 
After the excitation of the qubit with a $\pi$-pulse one would expect a probability $P\left(|1\rangle\right) = 1$ measuring the excited state.
In the experiment, the qubit already decayed by a certain degree after its preparation in the excited state until the time $t_0$ of the measurement.
The probability 
\begin{equation}
P \left (t_ {0}\,\right) = \text{e}^{- \frac{1}{T_{1}}t_ {0}} 
\end{equation}
to measure the excited state at a time $t_{0}$ therefore has a value between 1 and 0, depending on the relaxation rate $T_{1}^{-1}$ of the qubit to the ground state.
Each of the $N_{0}$ measurements are independent when the qubit is prepared again in the same state for each individual measurement.

\begin{figure}[htb]
	\includegraphics[width=\columnwidth]{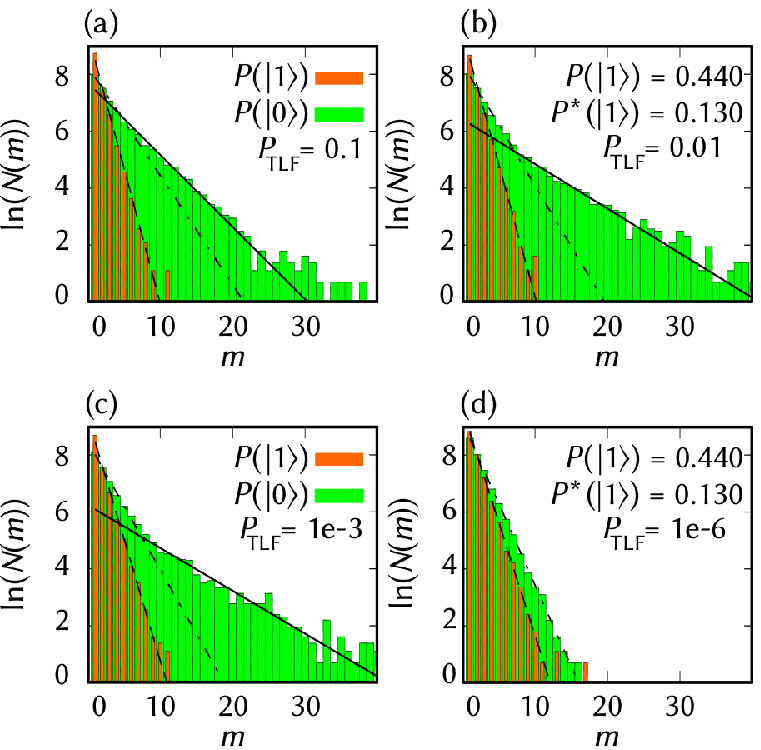}
         \caption{(color online) Numerical simulation of the abundance for
         several magnitudes of the TLS's switching probability $P_{\text{TLF}}$.
         The probabilities to measure the excited state of the qubit are $P\left(| 1\rangle \right) = 0.44$
         and $P^{\, *} \left (| 1 \rangle \right) = 0.13$.
         Different switching probabilities of the TLF 
         (a) $ P_{\text{TLF}} = 0.1 $ (b) $ P_{\text{TLF}} = 0.01 $ (c) $ P_{\text{TLF}} = 1 \cdot 10^{-3} $ (d) $ P_{\text{TLF}} = 1 \cdot 10^{-6} $
         make it clear that only rates within these magnitudes are resolvable with the used measurement protocol.
	}
	\label{fig:TS-statistics-sim}
\end{figure}

\smm{
Let $\mathcal{Q}$ be a set of the outcome of $N_{0}$ Bernoulli experiments.
We want to know how many subsets with exactly $m$ successes (further
denoted as \ket{1}) are in $\mathcal{Q}$. Therefore we look at the
probability of such a subset to contain exactly $m$ times \ket{1}.
In the next step we
expand by one element, which is \ket{1} with probability $p$. Finding
\ket{0} on the other hand determines one end of the chain with probability
$1-p$. Expanding in the other direction and finding \ket{0} again
determines the number of consecutive \ket{1} to be $m=0$. When finding
\ket{1} we go on to expand, where the probability of each step is
$p$ to find a \ket{1} and $1-p$ to find \ket{0}. If we find \ket{0}
we expand the other end, until it hits a \ket{0} as well.
}
\smm{For a chain with $m=0$ times \ket{1}, which corresponds to the probability
of two consecutive \ket{0} is $P(0)=(1-p)(1-p)$. For finding $m=1$
times \ket{1} which is separated from the rest by \ket{0}, being
exactly one consecutive \ket{1}, is $P(1)=(1-p)^{2}\cdot p$.
}

\smm{For arbitrary $m$ the probability of those minimal subsets that enclose
exactly $m$ times \ket{1} is
\begin{equation}
P(m)=(1-p)\cdot p^{m}\cdot(1-p)=(1-p)^{2}\cdot p^{m}.\label{eq:propability(m)}
\end{equation}}

\smm{We now want to know how many of such subsets of length $m$ are contained
in the set $\mathcal{Q}$ of length $N_{0}$. The closed expression of the abundance $N(m)=N_{0}P(m)$ taking the logarithm yields with equation (\ref{eq:propability(m)})
\begin{equation}
\log(N(m))=m\cdot\log(p)+\log(N_{0})+2\cdot\log(1-p).\label{eq:SM}
\end{equation}}

\smm{There are some subsets for which the probability is not $P(m)=(1-p)^{2}\cdot p^{m}$ but $P_{edge}(m)=(1-p)\cdot p^{m}$ as these two sets hit the border
of the set $\mathcal{Q}$ and do not need to end with two \ket{0}s.
As long as $m\ll N_{0}$, their number is small compared to the total
number of chains with length $m$ and in our case they can be neglected.
}

\smm{Finally we point out which timescales are resolvable with the simulation of coin tosses with two different biased coins modeling our correlated data.}
In Fig.~\ref{fig:TS-statistics-sim}(a)-(d) four numerically simulated experiments with different orders of magnitude of the switching probability of the TLF are presented to show the resolution of switching rates ranging from $1/\unit[0.7]{s}$ to $1/\unit[8]{ms}$, assuming a repetition rate of $\unit[1.2]{kHz}$.
The characteristic kink in the histograms is shifted to larger $m$ by lowering the fluctuation rate of the TLF.

\end{document}